\documentclass[aps,floats,twocolumn]{revtex4}
\usepackage[final]{graphicx}
\DeclareGraphicsExtensions{.eps,.ps,.pdf}
\usepackage{amsmath,amssymb,bbold,bm}
\def\id{\mathbb{1}}

\newcommand{\bk}{{\bf k}}

\newcommand{\br}{{\bf r}}

\newcommand{\ba}{{\bf a}}
\newcommand{\bd}{{\bf d}}

\newcommand{\bK}{{\bf K}}

\begin{document}

\title{Topological insulator on the kagome lattice}
\author{H.-M. Guo}
\affiliation{Department of Physics and Astronomy, University of British Columbia,Vancouver, BC, Canada V6T 1Z1}
\author{M. Franz}
\affiliation{Department of Physics and Astronomy, University of British Columbia,Vancouver, BC, Canada V6T 1Z1}
\date{\today}

\begin{abstract}
Itinerant electrons in a two-dimensional kagome lattice form a
Dirac semi-metal, similar to graphene. When lattice and spin symmetries
are broken by various periodic perturbations this semi-metal is shown
to spawn interesting non-magnetic insulating phases. These include a two-dimensional topological insulator with a non-trivial Z$_2$ invariant and robust gapless edge states, as well as dimerized and trimerized `Kekul\'{e}' insulators.  The latter two are topologically trivial
but the Kekul\'{e} phase possesses a complex order parameter with fractionally
charged vortex excitations. A charge density wave is shown to couple
to the Dirac fermions as an effective axial gauge field.
\end{abstract}

\maketitle

Certain physical observables in solids, such as the magnetic flux in
a superconductor or the Hall conductance in a quantum Hall liquid,
are precisely quantized despite the fact that the host material may
contain a significant amount of disorder. In all known cases this
quantization phenomenon can be attributed to the notion of
topological order. The bulk of such systems are characterized by a
topological invariant that is insensitive to microscopic details and
robust with respect to weak disorder. Recently, a new class of
topological invariants has been established to characterize all
time-reversal (${\cal T}$) invariant band insulators in 2 and 3
spatial dimensions \cite{kane2,mele1,moore1,roy1}. These new
invariants are of the Z$_2$ variety and the precisely quantized
physical observable is the number of gapless edge (surface) states
modulo 2. Topological insulators (TI) exhibit an odd number of edge
(surface) states while trivial insulators exhibit an even number,
possibly zero.  In many ways the edge (surface) states of a TI
behave as a perfect metal and are predicted to exhibit various
unusual properties \cite{qi1,essin1,qi2}. They also show promise as
possible components of future quantum computers \cite{fu1}.

Experimentally, HgTe/(Hg,Cd)Te quantum wells of certain width and
composition have been identified as 2D topological `spin Hall'
insulators with robust edge states \cite{konig1}. In addition, several
3D compounds involving bismuth have been so identified
\cite{cava1,cava2} and several more have been predicted
\cite{teo1,zhang1} as likely candidates. In view of these rapid
developments it appears likely that TIs might be a
fairly common occurrence in nature. Given their exotic properties and their
potential for technological applications it is important to identify
and study various  model systems that exhibit this behavior. Such
theoretical understanding will aid experimental searches for new
materials and help understand their unusual properties.

\begin{figure}[t]
\includegraphics[width=8.4cm]{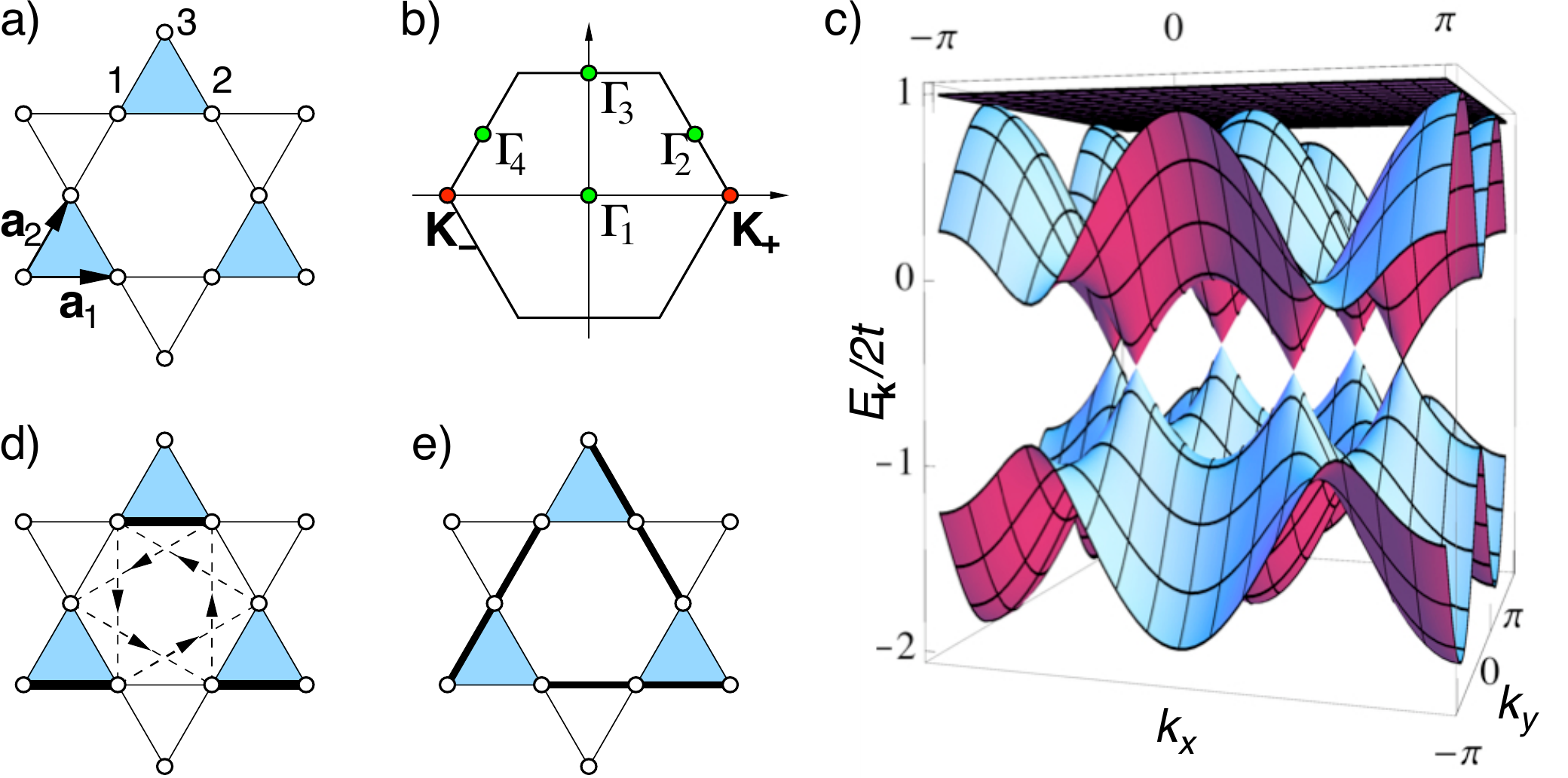}
\caption{(Color online) a) kagome lattice is a triangular Bravais
  lattice with a 3-point basis forming a shaded triangle. b) The first
  Brillouin zone with the nodal points $\bK_\pm$ and time-reversal
  invariant momenta ${\bm \Gamma}_n$ marked. c) The tight-binding band
  structure Eq.\ (\ref{ek0}). d) In the dimerized phase hopping amplitude
  along the thick (thin) bonds in $\ba_1$ direction is $t+\eta_1$
  ($t-\eta_1$). In the spin-orbit phase spin-up electrons hop between
  second neighbor sites with amplitude $i\lambda$ when moving along
  the arrow, $-i\lambda$ against the arrow. For spin-down electrons the
  arrows are reversed. e) Trimerized `Kekul\'{e}' phase.
}
\label{fig1}
\end{figure}

In this Brief Report we advance the above agenda by describing a new
class of 2-dimensional topological insulators on the kagome lattice,
Fig.\ \ref{fig1}a. Although the properties of spin systems on the
kagome lattice have been extensively studied, relatively little
attention has been paid to the non-magnetic insulating phases of
itinerant electrons. In what follows we demonstrate, both
analytically and numerically, that a simple tight-binding model of
electrons on the kagome lattice at both ${1\over 3}$ and ${2\over
3}$ filling becomes a topological insulator upon inclusion of the
spin-orbit (SO) coupling. Other ${\cal T}$-invariant insulating
phases include the dimerized and a trimerized band insulators; both
have trivial Z$_2$ invariants but the latter possesses a complex
order parameter with vortices that carry fractional charge, in
analogy to what happens in the Kekul\'{e} phase of graphene
\cite{chamon1}. We demonstrate that a charge density wave (CDW)
modulation of the on-site energies does not produce a spectral gap
but instead generates terms that couple as a U(1) gauge field to the
low-energy Dirac fermions at ${1\over 3}$ filling. Consistent with
${\cal T}$-invariance, the gauge field couples with opposite sign to
the two species of Dirac fermions, thus furnishing a concrete
realization of an {\em axial} gauge field in a solid state system.

We now proceed to substantiate the above claims. Our starting point is the tight-binding model
\begin{equation}\label{h0}
H_0=-t\sum_{\langle ij \rangle\sigma} c^\dagger_{i\sigma} c_{j\sigma},
\end{equation}
where $c^\dagger_{i\sigma}$ creates an electron with spin $\sigma$ on the site
$\br_i$ of the kagome lattice and $\langle ij \rangle$ denotes nearest neighbors. In momentum space Eq.\ (\ref{h0}) becomes $H_0=\sum_{\bk\sigma}\Psi^\dagger_{\bk\sigma}{\cal H}^0_\bk\Psi_{\bk\sigma}$ with
$\Psi_{\bk\sigma}=(c_{1\bk\sigma},c_{2\bk\sigma,}c_{3\bk\sigma})^T$ and
\begin{equation}\label{hk0}
{\cal H}^0_\bk=-2t
\begin{pmatrix}
  0 & \cos{k_1} & \cos{k_2} \\
  \cos{k_1} & 0 & \cos{k_3} \\
\cos{k_2} & \cos{k_3} & 0
\end{pmatrix}.
\end{equation}
The index $l=1,2,3$ in $c_{l\bk\sigma}$ labels the three basis sites in the
triangular unit cell; $\ba_1=\hat{x}$, $\ba_2=(\hat{x}+\sqrt{3}\hat{y})/2$ and
$\ba_3=\ba_2-\ba_1$ denote the three nearest neighbor vectors, and
 $k_n=\bk\cdot\ba_n$.

The spectrum of ${\cal H}^0_\bk$, Fig.\ \ref{fig1}c, consists of one flat band $E_{\bk}^{(3)}=2t$ and two dispersive bands
\begin{equation}\label{ek0}
E_\bk^{(1,2)}=t\left[-1\pm\sqrt{4A_\bk-3}\right],
\end{equation}
with $A_\bk=\cos^2{k_1}+\cos^2{k_2}+\cos^2{k_3}$. Bands 1 and 2 touch
at two inequivalent Dirac points  $\bK_{\pm}=(\pm 2\pi/3,0)$ located at the
corners of the hexagonal Brillouin zone (BZ), Fig.\ \ref{fig1}b. Bands 2 and 3
 touch at the center of the BZ.

At ${1\over 3}$ filing the lowest band is filled and the low-energy
electronic excitations of $H_0$ resemble those of graphene. To produce
an insulator we now seek terms bi-linear in the electron operators
that lead to the formation of a gap at the Dirac points. First, we
focus on perturbations that do not further break the translational
symmetry of $H_0$ and preserve ${\cal T}$. We have been able to
identify two such terms: (i) a spin-independent lattice dimerization and
(ii) a spin-orbit interaction induced hopping between the second neighbors
\cite{haldane1,kane1,kane2}. The former term breaks the inversion
symmetry of the lattice while the latter breaks the SU(2) spin
symmetry.

The dimerization is described by
$H_{\rm dim}=-\sum_{\langle ij \rangle\sigma} \delta t_{ij}c^\dagger_{i\sigma} c_{j\sigma},$
where $\delta t_{ij}=\pm \eta_n$ describes an alternating pattern of
bond hopping integrals along the three principal spatial directions as
illustrated in Fig.\ \ref{fig1}d. In $k$-space this becomes

\begin{equation}\label{hkdim}
{\cal H}^{\rm dim}_\bk=2
\begin{pmatrix}
  0 &i\eta_1 \sin{k_1} & i\eta_2\sin{k_2} \\
   & 0 & i\eta_3\sin{k_3} \\
   &   & 0
\end{pmatrix},
\end{equation}
for both spin projections. (The lower triangle of the matrix is
understood to be filled so that the matrix is hermitian.)  The full
expression for the spectrum of ${\cal H}^0_\bk+{\cal H}^{\rm dim}_\bk$
is complicated but it is easy to see that a gap $\Delta_{\rm
dim}\simeq 2|\eta|$ with $\eta= \eta_1+\eta_2+\eta_3$ exists at the Dirac
points.

The spin-orbit term takes form
\begin{equation}\label{hso}
H_{\rm SO}=i{2\lambda\over\sqrt{3}}\sum_{\langle\langle ij \rangle\rangle\alpha\beta}
 (\bd_{ij}^1\times\bd_{ij}^2)\cdot{\bm \sigma}_{\alpha\beta} c^\dagger_{i\alpha} c_{j\beta},
\end{equation}
where $\lambda$ is the spin-orbit coupling strength, $\bd_{ij}^{1,2}$
are nearest neighbor vectors traversed between second neighbors $i$
and $j$, and ${\bm \sigma}$ is the vector of Pauli spin matrices. Since
$\bd_{ij}^{1,2}$ all lie in the $xy$ plane in our 2D model, only
$\sigma_3$ appears in Eq.\ (\ref{hso}) and the Hamiltonian decouples
for the two spin projections along the $z$ axis. The pattern of spin-orbit
induced second neighbor hoppings then resembles the Haldane model
\cite{haldane1} and is illustrated in Fig.\ \ref{fig1}d. In $k$-space one
obtains
\begin{equation}\label{hkso}
{\cal H}^{\rm SO}_\bk=\pm 2\lambda
\begin{pmatrix}
  0 &i\cos{(k_2+k_3)} & -i\cos{(k_3-k_1)} \\
   & 0 & i\cos{(k_1+k_2)} \\
   &   & 0
\end{pmatrix},
\end{equation}
where the $+(-)$ sign refers to spin up (down) electrons. Once again,
although the full spectrum is complicated it is easy to deduce that a
gap $\Delta_{\rm SO}=4\sqrt{3}|\lambda|$ opens up at the Dirac points.
We remark that $H_{\rm SO}$ opens a gap also between bands 2 and 3 while
$H_{\rm dim}$ does not.

In order to develop some intuition for these insulating phases it is
useful to examine the form of the low-energy Hamiltonians governing
the excitations in the vicinity of the two Dirac points.  This is
obtained by linearizing ${\cal H}_\bk={\cal H}^0_\bk+{\cal H}^{\rm
dim}_\bk+{\cal H}^{\rm SO}_\bk$ near $\bK_\pm$ and subsequently
projecting onto the subspace associated with bands 1 and 2. Assuming
that both $\eta$ and $\lambda$ are non-zero we obtain 4 independent
Dirac Hamiltonians,
\begin{equation}\label{hlin}
h_{\bk \ell\alpha}=v(\tau_3k_x+\tau_1 k_y)+\tau_2m_{\ell\alpha},
\end{equation}
labeled by the spin index $\alpha=\pm$ and `valley' index $\ell=\pm$.
$\tau_j$ are Pauli matrices acting in the space spanned by the
degenerate eigenstates of ${\cal H}^0_{\bK_+}$ and ${\cal H}^0_{\bK_-}$,
$v=\sqrt{3}t$ is the Fermi velocity and
\begin{equation}\label{hlin_m}
m_{l\alpha}=2\sqrt{3}\alpha\lambda+\ell\eta
\end{equation}
are Dirac masses whose relative signs define two distinct phases of
the system.  When $|\eta|> 2\sqrt{3}|\lambda|$ dimerization dominates and
the Dirac masses at $\bK_+$ and $\bK_-$ exhibit opposite signs, independent
of spin. When $|\eta|< 2\sqrt{3}|\lambda|$ the SO interaction dominates and the
mass signs at the two Dirac points are the same for a given spin but differ
for the opposite spin projections. The two phases meet at a pair of
critical lines $|\eta|= 2\sqrt{3}|\lambda|$. When crossing these lines, two
out of four gaps close and the associated Dirac masses change signs.

To understand the significance of the mass signs consider a boundary
between the two phases, running along, say, the $x=0$ line in real
space. For concreteness and simplicity we take $\lambda>0$, $\eta=0$ in
the left half-plane and $\lambda=0$, $\eta>0$ in the right
half-plane. Focusing first on the Dirac point $\bK_+$, we note that the
spin-up mass $m_{++}$
remains positive for all $x$, suggesting a fully gapped spectrum
everywhere. The spin-down mass, however, necessarily
undergoes a sign change across the $x=0$ boundary. Such a soliton mass
profile is known to produce massless states \cite{jackiw1} in the
associated Dirac equation, localized near the boundary. Specifically,
Dirac equation
\begin{equation}\label{dir1}
[v(-i\tau_3\partial_x-i\tau_1 \partial_y)+\tau_2m(x)]\phi(x,y)=E\phi(x,y)
\end{equation}
with $m(x\to\pm\infty)=\pm m_0$ has a gapless solution
\begin{equation}\label{dir2}
\phi_k(x,y)=
\begin{pmatrix}
1 \\ 1
\end{pmatrix}
e^{iky}e^{-{1\over v}\int_0^xm(x')dx'},
\end{equation}
extended along the boundary but localized in the transverse direction,
with linearly dispersing energy $E_k=vk$. Similar analysis leads to a gapless
 state at node $\bK_-$, but now for spin up and with $E_k=-vk$.

A pair of spin-filtered, oppositely dispersing gapless edge states  is
a hallmark of the topological spin-Hall insulator
\cite{kane2,mele1,moore1}. The fact, apparent from the above construction,
that these edge states depend only on the bulk band structure of the
two insulators  and not on the details of the edge indicates their
topological origin. It follows that {\em one} of these phases must be
a topological insulator. It is easy to see that the dimerized phase is
smoothly connected to a trivial insulator. Consider increasing all
$\eta$'s continuously until $\eta_1=\eta_2=\eta_3=t$. At this point
the kagome lattice breaks down into a collection of disconnected
elementary triangles. This, clearly, is a trivial insulator. The
spin-orbit phase, on the other hand, cannot be smoothly deformed into
a trivial insulator and we show below by an explicit calculation that
it indeed possesses a nontrivial Z$_2$ invariant \cite{kane2,mele1,moore1}.

When a crystal possesses inversion symmetry the Z$_2$ topological
invariant $\nu$ is easy to evaluate. According to Ref.\ \cite{fu2} $\nu$ is
related to the parity eigenvalues $\xi_{2m}({\bm \Gamma}_i)$ of the
2$m$-th occupied energy band at the four ${\cal T}$-invariant momenta
${\bm \Gamma_i}$. Our system is inversion symmetric when all $\eta$'s vanish
and so we can use this method to find $\nu$. If we select site 1 of the
unit cell as the center of inversion then the parity operator acts as
${\cal P}[\psi_1(\br),\psi_2(\br),\psi_3(\br)]=
[\psi_1(-\br),\psi_2(-\br-2\ba_1),\psi_3(-\br-2\ba_2)]$ on the triad
of the electron wavefunctions in the unit cell labeled by vector
$\br$. In momentum space the parity operator becomes a diagonal
$3\times 3$ matrix ${\cal P}_\bk= {\rm diag}(1,e^{-2i\ba_1\cdot
\bk},e^{-2i\ba_2\cdot \bk})$. The four ${\cal T}$-invariant momenta in
our system are marked in Fig.\ \ref{fig1}b and can be expressed as
${\bm \Gamma}_i=\pi(\hat{x}+\hat{y}/\sqrt{3})n_i/2+\pi(-\hat{x}+\hat{y}/\sqrt{3})m_i/2$
with $n_i,m_i=0,1$.  It is straightforward to obtain the eigenstates
of ${\cal H}_{{\bm \Gamma}_i}$ numerically and determine the parity
eigenvalues of the occupied bands. We find that three $\xi$'s are
positive and one is negative.  Which of the four $\xi$'s is negative
depends on the choice of the inversion center but the product
$\Pi_i\xi({\bm \Gamma}_i)=(-1)^\nu$ is independent of this choice and
determines the non-trivial Z$_2$ invariant $\nu=1$, confirming our
hypothesis that the spin-orbit phase at ${1\over 3}$ filling is a
topological insulator. Similar considerations for
${2\over 3}$ filling also yield $\nu=1$.

When the dimerization is present the inversion symmetry is broken and
we must use the more general method of Ref.\ \cite{kane2} to find
$\nu$. This relies on counting the number of pairs of first order
zeros of the quantity $P(\bk)={\rm Pf}\langle
u_{m\bk}|\Theta| u_{n\bk}\rangle$, where $u_{m\bk}$ is the $m$-th
eigenstate of ${\cal H}_\bk$, $\Theta$ is the time-reversal operator
and the Pfaffian is taken over occupied bands $m$, $n$. We obtain $u_{m\bk}$ by
the numerical diagonalization of ${\cal H}_\bk$ and use it to
straightforwardly evaluate $P(\bk)$. We find that the latter contains
exactly one pair of first order zeros (located at $\bK_\pm$) in the SO
phase when  $|\eta|< 2\sqrt{3}|\lambda|$, indicating $\nu=1$. When  $|\eta|>
2\sqrt{3}|\lambda|$ the zeros disappear and the system becomes a
trivial insulator with $\nu=0$.

\begin{figure}[t]
\includegraphics[width=8cm]{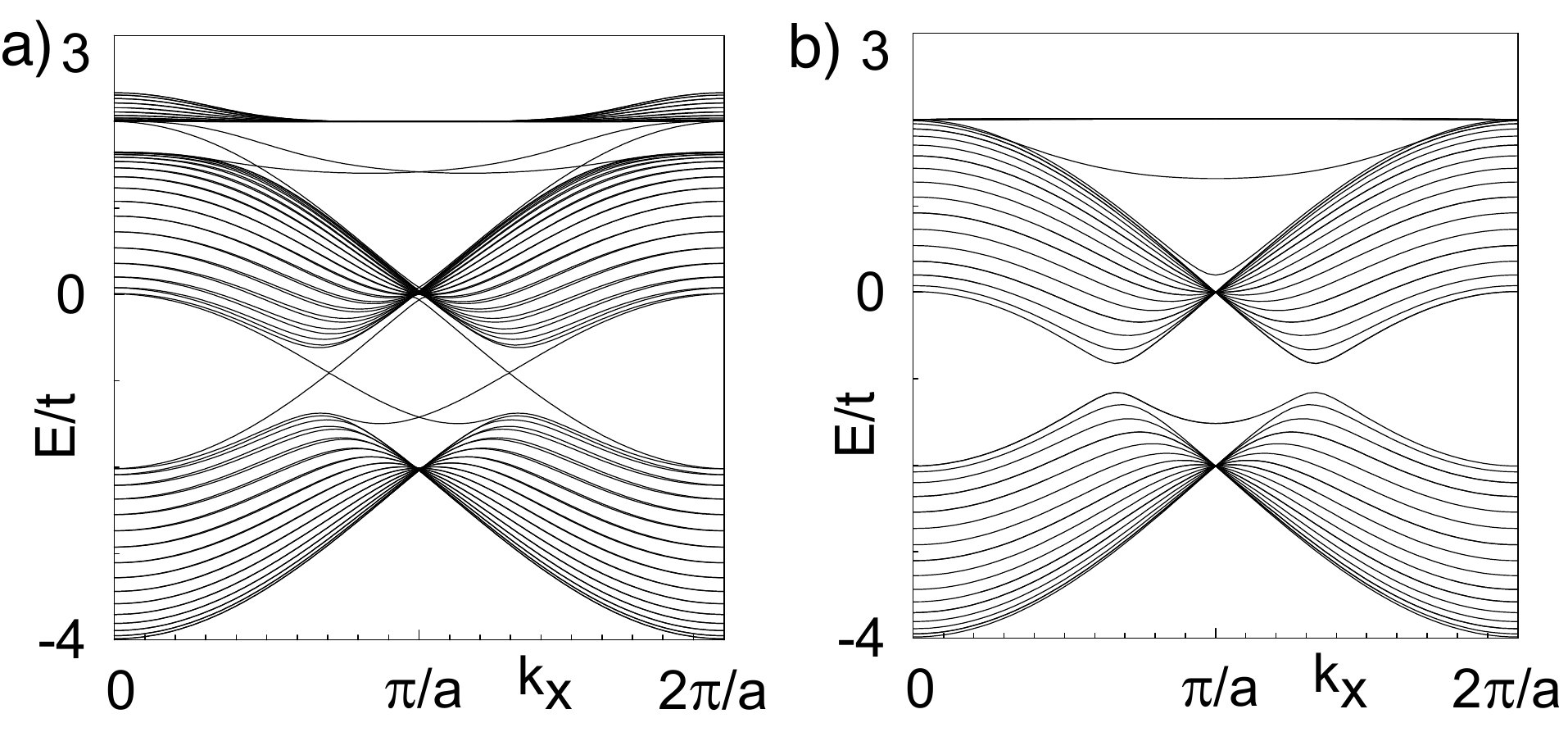}
\caption{Edge states in the lattice model for (a) spin-orbit and (b)
 dimerized insulator. A strip of width $N_y=16$ unit cells with
 open boundary conditions along $y$ and infinite along $x$ is used with
 $\lambda/t=0.1$ and $\eta_1/t=0.1$ for (a) and (b), respectively. }
\label{fig2}
\end{figure}
To further support our identification of the SO phase as a
topological insulator we have performed numerical diagonalizations of
the lattice Hamiltonians $H_0+H_{\rm dim}$ and $H_0+H_{\rm SO}$ using the
strip geometry. In accord with the above arguments we find a
pair of robust spin-filtered gapless states associated with each edge
in the SO phase, Fig.\ \ref{fig2}a, traversing the gap between bands 1
and 2. A similar pair of states traverses the gap between bands 2 and
3, confirming that the SO phase is a topological insulator at ${1\over
  3}$ and ${2\over 3}$ filling.  In the dimerized phase there are
generically no such robust edge states, Fig.\ \ref{fig2}b, although for
certain values of parameters and types of edges gapless edge states
can occur. The latter mirror the edge states found along certain
types of edges in graphene \cite{nakada1} but are not topological in character.

In graphene, a staggered on-site potential is known to open up a gap at the
 Dirac points \cite{semenoff1}. We thus investigate the effect of the
 analogous CDW term,
\begin{equation}\label{hkcdw}
{\cal H}^{\rm CDW}_\bk={\rm diag}(\mu_1,\mu_2,\mu_3),
\end{equation}
on the kagome semi-metal, where $\mu_l$ represent the on-site potentials
of the $l=1,2,3$ basis sites of the lattice, independent of
spin. Repeating the procedure leading to Eq.\ (\ref{hlin}) we find the
low-energy Dirac Hamiltonians
\begin{equation}\label{hlin2}
h_{\bk \ell\alpha}=v\left[\tau_3(k_x-{\cal A}_x^{\ell})+\tau_1(k_y-{\cal A}_y^{\ell})\right]+\id\mu,
\end{equation}
for node $\ell$ and spin $\alpha$, with $\mu=(\mu_1+\mu_2+\mu_3)/3$ and
\begin{eqnarray}\label{gauge}
 {\cal A}_x^{\ell}&=&(2\mu_3-\mu_1-\mu_2)\ell/6\sqrt{3}t, \nonumber \\
{\cal A}_y^{\ell}&=&(\mu_1-\mu_2)\ell/6t.
\end{eqnarray}
The CDW couples to the Dirac fermions as a {\em gauge
field}. Owing to the factor $\ell$ in Eq.\ (\ref{gauge}), the gauge potential has the opposite sign at the two Dirac points. Consequently,  ${\cal A}$
must be thought of as an {\em axial} gauge field. This is consistent
with the fact that the CDW does not break ${\cal T}$.

If we allow for additional fermion bi-linears that break the translational
symmetry of $H_0$ then many other insulating phases become
possible. Of these, we briefly mention but one that realizes the
analog of the Kekul\'{e} phase in graphene \cite{chamon1}. This occurs for
a perturbation with a wavevector spanning the two nodal points $\bK_+$
and $\bK_-$ in Fig.\ \ref{fig1}b. In the linearized theory such a
perturbation leads to a Dirac mass that is off-diagonal in the space
of nodal Hamiltonians and is in general complex-valued. In graphene,
vortices in such a complex mass are known to carry a fractional charge
$\pm e/2$ (assuming spinless electrons) \cite{chamon1} and obey
fractional exchange statistics \cite{seradjeh1}. On the kagome
lattice one possible realization of such a perturbation is the
`trimerization' depicted in Fig.\ \ref{fig1}e. The unit cell contains
9 atoms and there are 3 distinct degenerate ground states related to
the one depicted by translations through $2\ba_1$ and $2\ba_2$. We
have verified that this perturbation indeed opens up a gap at the
Dirac points and the three patterns produce off-diagonal Dirac masses
with different complex phases. In analogy with the domain wall in the
trimerized 1D chain \cite{schrieffer1} we expect the vortex in the
complex mass to bind fractional charge $\pm e/3$, $\pm 2e/3$. This is
also suggested by the Goldstone-Wilczek type counting argument
\cite{goldstone1} generalized to 2 spatial dimensions
\cite{seradjeh2} that can be constructed for the kagome lattice
 \cite{franz_un}.  We leave the details of this and other interesting
related problems, such as the exchange statistics of these objects,
to future study.

A question naturally arises as to the experimental realization of
a system of itinerant electrons on the kagome lattice that would support
the exotic phenomena predicted above.
Copper atoms in ZnCu$_3$(OH)$_6$Cl$_2$, known as Herbertsmithite
\cite{helton1}, and iron atoms in Jarosites \cite{wills1}, provide good model Heisenberg kagome antiferromagnets. The kagome lattice has also been argued to play a role in transport and magnetic properties of layered
cobalt oxides, such as Na$_x$CoO$_2$ \cite{maekava1}. We hope that our
theoretical findings will provide motivation for future experimental searches
into possible realizations of the kagome lattice with itinerant
electrons close to ${1\over 3}$ or ${2\over 3}$ filling. We note that
it might be possible to artificially engineer Hamiltonian
(\ref{h0}) by modulating the two-dimensional electron gas with a
periodic potential with kagome symmetry, as recently demonstrated
for `artificial graphene' \cite{west1}.

When the basic tight-binding
Hamiltonian is at hand the additional terms required to form an
insulator can come about in various ways. In a crystal the SO coupling
arises naturally. In graphene the relevant coupling strength is too
small to open up a significant gap but in a lattice made of heavier
ions $\lambda$ will be larger. The perturbations considered above can also
arise as interaction-driven instabilities of itinerant electrons on
the lattice. For the Dirac semi-metal at ${1\over 3}$ filling a finite
interaction strength $U$ (of order $t$) is needed to open up a gap
\cite{raghu1}. At ${2\over 3}$ filling, however, the band crossing is
quadratic and the instability towards the insulating phase occurs at
infinitesimal (repulsive) interaction \cite{sun1}, making such system
a promising candidate for a 2D topological insulator.

\emph{Acknowledgment}.--- Authors are indebted to T. Aitken,
G. Rosenberg, B. Seradjeh and C. Weeks for stimulating
discussions. Support for this work came from NSERC, CIfAR and The
China Scholarship Council.


\begin{thebibliography}{10}

\bibitem{kane2} C.~L. Kane, and E.~J. Mele, \prl {\bf 95}, 146802 (2005).

\bibitem{mele1}
L.~Fu, C.~L. Kane, and E.~J. Mele, \prl {\bf 98} 106803 (2007).

\bibitem{moore1} J.~E. Moore and L.~Balents, \prb {\bf 75} 121306(R) (2007).

\bibitem{roy1} R.~Roy, \prb {\bf 79}, 195322 (2009).

\bibitem{qi1}
X.-L.~Qi, T.L.~Hughes, and S.-C.~Zhang, \prb {\bf 78}, 195424
(2008).

\bibitem{essin1}
A.M. Essin, J.E. Moore, D. Vanderbilt, \prl {\bf 102}, 146805
(2008).

\bibitem{qi2}
X.-L.~Qi, R.~Li, J.~Zang, S.-C.~Zhang, Science {\bf 323}, 1184
(2009).

\bibitem{fu1}
L.~Fu and C.~L. Kane, \prl {\bf 100}, 096407 (2008).

\bibitem{konig1} M. K\"{o}nig {\em et al.}, Science {\bf 318}, 766 (2007).

\bibitem{cava1} D.~Hsieh {\em et al.},
Nature {\bf 452}, 970 (2008).

\bibitem{cava2}
Y.~Xia {\em et al.},
arXiv:0812.2078.

\bibitem{teo1}
J.~C.~Y. Teo, L.~Fu, and C.~L. Kane, \prb {\bf 78} 045426 (2008).

\bibitem{zhang1}
H.-J.~Zhang {\em et al.},
\prb {\bf 80} 085307 (2009).

\bibitem{chamon1}
C.-Y.~Hou, C.~Chamon, and C.~Mudry, \prl {\bf 98}, 186809 (2007).

\bibitem{haldane1} F.D.M. Haldane, \prl {\bf 61}, 2015 (1988).

\bibitem{kane1} C.~L. Kane, and E.~J. Mele, \prl {\bf 95}, 226801 (2005).

\bibitem{jackiw1}
R.~Jackiw and C.~Rebbi, \prd {\bf 13}, 3398 (1976).

\bibitem{fu2} L.~Fu and C.~L. Kane, \prb {\bf 76}, 045302 (2007).

\bibitem{nakada1} K.~Nakada, M.~Fujita, G.~Dresselhaus and  M.S.~Dresselhaus, \prb {\bf 54}, 17954  (1996).

\bibitem{semenoff1} G.W.~Semenoff, \prl {\bf 53}, 2449 (1984).


\bibitem{seradjeh1}
B.~Seradjeh and M.~Franz, \prl {\bf 101}, 146401 (2008).

\bibitem{schrieffer1} W.P.~Su and J.R.~Schrieffer, \prl {\bf 46}, 738 (1981).

\bibitem{goldstone1}
J.~Goldstone and F.~Wilczek, \prl {\bf 47}, 986 (1981).

\bibitem{seradjeh2}
B.~Seradjeh, C.~Weeks and  M.~Franz, \prb {\bf 77}, 033104 (2008).

\bibitem{franz_un} M. Franz, (unpublished).

\bibitem{helton1} J.S.~Helton, {\em et al.} \prl {\bf 98}, 107204 (2007).

\bibitem{wills1} A.S.~Wills, A.~Harrison, C.~Ritter and R.I.~Smith,
 \prb {\bf 61}, 6156 (2000).

\bibitem{maekava1} W.~Koshibae and S.~Maekawa, \prl {\bf 91}, 257003 (2003).

\bibitem{west1} M.~Gibertini,  {\em et al.}, \prb {\bf 79}, 241406(R) (2009).

\bibitem{raghu1} S.~Raghu, X.-L.~Qi, C.~Honerkamp and S.-C.~Zhang, \prl {\bf 100}, 156401 (2008).

\bibitem{sun1} K.~Sun, H.~Yao, E.~Fradkin and S.A.~Kivelson, \prl {\bf 103}, 046811 (2009).



\end{thebibliography}

\end{document}